# GENERIC APPROACH TO VISUALIZATION OF TIME SERIES DATA

Sathya Krishnan TS
Department Computer Science and Engineering
Puducherry Technological University
Puducherry, India
satyakrishnan.s@pec.edu

Dr.P. Shunmugapriya
Department of Computer Science
Pondicherry University
Puducherry, India
pshunmugapriya@gmail.com

***Abstract* - Time series is a collection of data instances that are ordered according to a time stamp. Stock prices, temperature, etc are examples of time series data in real life. Time series data are used for forecasting sales, predicting trends. Visualization is the process of visually representing data or the relationship between features of a data either in a two-dimensional plot or a three-dimensional plot. Visualizing the time series data constitutes an important part of the process for working with a time series dataset. Visualizing the data not only helps in the modelling process but it can also be used to identify trends and features that cause those trends. In this work, we take a real-life time series dataset and analyse how the target feature relates to other features of the dataset through visualization. From the work that has been carried out, we present an effective method of visualization for time series data which will be much useful for machine learning modelling with such datasets.**

## I. INTRODUCTION

Time Series has been around for a long time, and it is one of the important areas where the application of Machine Learning (ML) has been huge. ML and deep learning models, work on time series data to predict or forecast the target variable of the historical data fed to the model. Stock market companies [9], e-commerce companies like Amazon have largely benefitted by the use of ML by predicting which company to invest in, which product to promote on what time of the year and many other strategies that can be extracted from the time series data they have. A good ML model is required for such important tasks and visualization is an important part for building such a model. Visualization is the process of transferring the given data onto a plot or graph either in 2-dimension or in 3-dimension that can be easily understood by a non-technical person. Visualization is a tool that is used to understand the data better and the relations between the features present that is inconspicuous in a tabular format [10]. In time series, visualization is used a lot to weed out the unnecessary features, find features causing different type of trends and find features that affect the target feature the most [11]. In this paper, we present the visualization work done on a time series dataset from a Kaggle competition called 'Tabular Playground Series – May 2022". Data preparation and feature engineering [2] performed on the dataset have been explained in the following sections. From the figure presented and the various processes used to develop those plots, standard steps are selected to develop a generalized method for visualizing time series data. The method presented can be used to develop an understanding of the relations that a target feature has with the other features in the data and also the common trends present in the data.

## II. RELATED WORK

In Machine Learning and Time Series, visualization is used extensively to arrive at conclusions and to aid in the modelling process. For Machine Learning applications, the process of visualization has been well documented and researched but when it comes to Time Series, works related to visualization are very few.

Yujie Fang et al., in their work [6] have compiled and analysed the various types of visualization for time series data. Their work also illustrates what type of visualization to use for a particular problem and how to use them. But it does not propose a general method for time series data visualization, one that anyone starting with time series can use. In this work, we have taken on this problem and have come up with a generic method that can be used as a starting point for time series data visualization, one that anyone starting with time series visualization can employ.

## III. TIME SERIES

Time series is a collection of data instances that are ordered according to a time stamp, or it is a measurement of an item over a period of time. For example, the price of oil for the last three months represents a time series where the price of oil is the 'item' in definition, measured during each day for the last three months. Time Series are widely used in econometrics, weather and price forecasting, etc. It is made up of four components – Trend, Seasonal Variations, Cyclic Variations and Random or Irregular movements. Trend is also called as Long-Term movement whereas Seasonal and Cyclic variations are called as Short-Term movements.

Trend represents the general increase or decrease of the quantity over the period for which the data was collected. Trend represents the central tendency of a time series, and it can either be upward, downward, or stable.

Seasonal variations represent the variations in the values of the quantity for a particular period and it is likely to happen the next time the period arrives. Temperature of a city recorded over the years, follows seasonal variation.

Cyclic variations represent the variations that are recurrent and last for more than a year. They follow a well determined pattern. For example, a Business cycle consists of prosperity, recession, depression and recovery.

Random or irregular movements account for the things that we cannot predict. These movements are unforeseen and erratic and most of the time they are not taken into account while analysing the time series data. For example, the occurrence of a natural disaster altering the trend component of the time series.

## IV. TIME SERIES VISUALIZATION

Time series visualization is the process of plotting out the variations of the target feature with respect to some other feature, over some period of time. The period of time may be measured in months, days and hours depending on the target feature. For example, the visual representation of change in stock prices over the last 10 days. Visualization plays an important role when it comes to time series problems as the field of time series requires a lot of visual representation to simplify the statistical analysis done on the data. It also presents the components of a time series in an understandable way. Visualization will be of great help in selecting the features that are most useful for the problem at hand. It picks out these features by analysing the variations of the target feature with the other features. It is also helpful in building machine learning and deep learning models related to time series problems. With these points in mind, we have developed a generic and efficient method made up of 5 steps that can be used as an entry point for visualizing time series data.

## GENERIC APPROACH TO VISUALIZE TIME SERIES DATA

This section proposes the method that has been developed from the visualizations shown in the next section. This method is a product of the experiments carried out while trying to visualize the dataset used for this paper and the method proposed here is a general method that can be used as an entry point for visualizing most of the time series data.

1. First and foremost thing, is to select the unit of measurement like hour, day, month that suits the target variable and gives us a lot of scope for analysis.
2. After selecting the unit of measurement, create a plot with the unit of measurement along x-axis and the target variable along the y-axis.
3. Use the categorical features in the dataset as 'hue' parameter to split the plot mentioned in step 2. If the plot becomes crowded split the 'hue' parameter into groups and analyse them. The purpose of splitting the plot is to know how the target variable varies during a particular period according to the 'hue' parameter.

   Hue is a feature from the dataset made up of categorical values that groups the points in the plot into the respective categories.
4. For the continuous valued features, split them into bins using Sturge's rule,

   $$Number\ of\ bins = 1 + log_2 N$$

   where N is the number of instances. Then use those bins as 'hue' parameter to split the plot. This division of continuous valued columns will also showcase the trends in the target variable.

   Sturge's law is used whenever the number of bins in a histogram plot is high. It helps in reducing the number of bins which in turn helps in visualization as there are less number of categories and the meaning that the plot is trying to convey can be understood easily.
5. After completing step 3 and 4, pick the features that form the most distinct classes in the plot and try to use the combination of these features as the 'hue' parameter to split the plot further.

## V. TIME SERIES VISUALIZATION OF THE KAGGLE DATASET:

In order to visualize a dataset, certain processes have to be carried out on the dataset before it can be visualized. Each process is explained briefly below. First the dataset must be cleaned and imputed. The second step is 'feature engineering' where new features are derived or extracted from the available features. In this step the features can also be transformed into a form that will be suitable for visualization. After the above processes have been completed, the dataset can be visualized. The following sections show how the above mentioned steps have been carried out on the Kaggle dataset.

Step 1: Clean and Impute

The dataset must be cleaned and imputed. Cleaned, here means that values belonging to a certain category must be of the same datatype and imputed means that there should be no 'NULL' values in the dataset. The presence of 'NULL' values will affect the statistical values of the data like average, standard deviation.

Step 2: Feature Engineering

Data in its original form can be used for visualization but most of the times there will be some important data hidden among the features and those features have to extracted from the available features, for example, the *time* feature in the dataset comprises of the date, hour, minute and second. Important information like what day it is, whether a particular day is a weekday or weekend, etc can be extracted from it. Data should be cleaned, imputed, and converted to a form preferred to the problem [4].

Step 3: Visualize

Once the above steps have been carried out, the data is ready to be visualized. Visualization will be carried out by the method proposed.

### *A. DATASET DESCRIPTION*

The dataset that has been used for this work, is taken from a Kaggle competition called "Tabular Playground Series – Mar 2022". The dataset used in the competition has been derived from Chicago Traffic Tracker – Historical Congestion Estimates dataset [3]. The dataset consists of the measurement of traffic congestion across 65 roadways from April through September of 1991. It is made up of 848835 instances and five features. Fig. 1. shows the first few instances of the dataset.

```
df.head()
```

| row_id | time | x | y | direction | congestion |
|---|---|---|---|---|---|
| 0 | 1991-04-01 00:00:00 | 0 | 0 | EB | 70 |
| 1 | 1991-04-01 00:00:00 | 0 | 0 | NB | 49 |
| 2 | 1991-04-01 00:00:00 | 0 | 0 | SB | 24 |
| 3 | 1991-04-01 00:00:00 | 0 | 1 | EB | 18 |
| 4 | 1991-04-01 00:00:00 | 0 | 1 | NB | 60 |

Fig. 1. Sample instances from the dataset

The dataset consists of no 'NULL' values. Here the *row_id* is a unique identifier for each instance, time is the 20-minute period in which the measurement was taken and it is of type 'string', *x* is the east-west midpoint coordinate of the roadway, *y* is the north-south midpoint coordinate of the roadway, *direction* is the direction of travel and finally the target variable *congestion* represents the congestion levels for the roadways during each hour. The values for congestion have been normalized to range from 0 to 100.

## B. FEATURE ENGINEERING

As shown in Fig. 1. the data in its original form cannot be used for visualization as the year, month, date and time values have been put together in a string. Those values have to be separated and new columns have to be created for each of the above-mentioned features. First the date attributes like month, date and year have been separated from the 'time' attribute and they have been added as separate features as presented in Fig. 2. Next, the time related attributes like hour, minute and second have derived from the 'time' attribute and they have been added as features to the dataset.

| row_id | time | x | y | direction | congestion | date | year | month | hour | minute | second |
|---|---|---|---|---|---|---|---|---|---|---|---|
| 0 | 00:00:00 | 0 | 0 | EB | 70 | 1 | 1991 | 4 | 0 | 0 | 0 |
| 1 | 00:00:00 | 0 | 0 | NB | 49 | 1 | 1991 | 4 | 0 | 0 | 0 |
| 2 | 00:00:00 | 0 | 0 | SB | 24 | 1 | 1991 | 4 | 0 | 0 | 0 |
| 3 | 00:00:00 | 0 | 1 | EB | 18 | 1 | 1991 | 4 | 0 | 0 | 0 |
| 4 | 00:00:00 | 0 | 1 | NB | 60 | 1 | 1991 | 4 | 0 | 0 | 0 |

Fig. 2. Dataset form after feature engineering on time feature

After forming these attributes from 'time' feature, we can then extract some more information like weekday or weekend, day of a particular instance and many other details. Extracting as many details as possible without data redundancy is required. The final form of the dataset before using it for data visualization is shown in Fig. 3.

| row_id | x | y | direction | congestion | date | year | month | hour | minute | weekday | day | PM |
|---|---|---|---|---|---|---|---|---|---|---|---|---|
| 175256 | 0 | 3 | WB | 45 | 8 | 1991 | 5 | 15 | 20 | 1 | Wednesday | 1 |
| 524922 | 2 | 1 | SB | 37 | 22 | 1991 | 7 | 11 | 20 | 1 | Monday | 0 |
| 651648 | 1 | 1 | EB | 46 | 18 | 1991 | 8 | 15 | 20 | 0 | Sunday | 1 |
| 680838 | 1 | 2 | NB | 59 | 24 | 1991 | 8 | 21 | 40 | 0 | Saturday | 1 |
| 136445 | 0 | 2 | WB | 38 | 30 | 1991 | 4 | 6 | 20 | 1 | Tuesday | 0 |

Fig. 3. Final form of the dataset

## C. EXPERIMENT AND RESULTS

This section of the work shows how the combination of features can be used to represent the variation of congestion over a period. First, a simple plot is shown. The variation of congestion with respect to month is shown in Fig. 4. We can infer that till June, the traffic flow was following a decreasing trend, but it suddenly increased for the month of August indicating that during the month of August there might have been some important event or festival in Chicago.

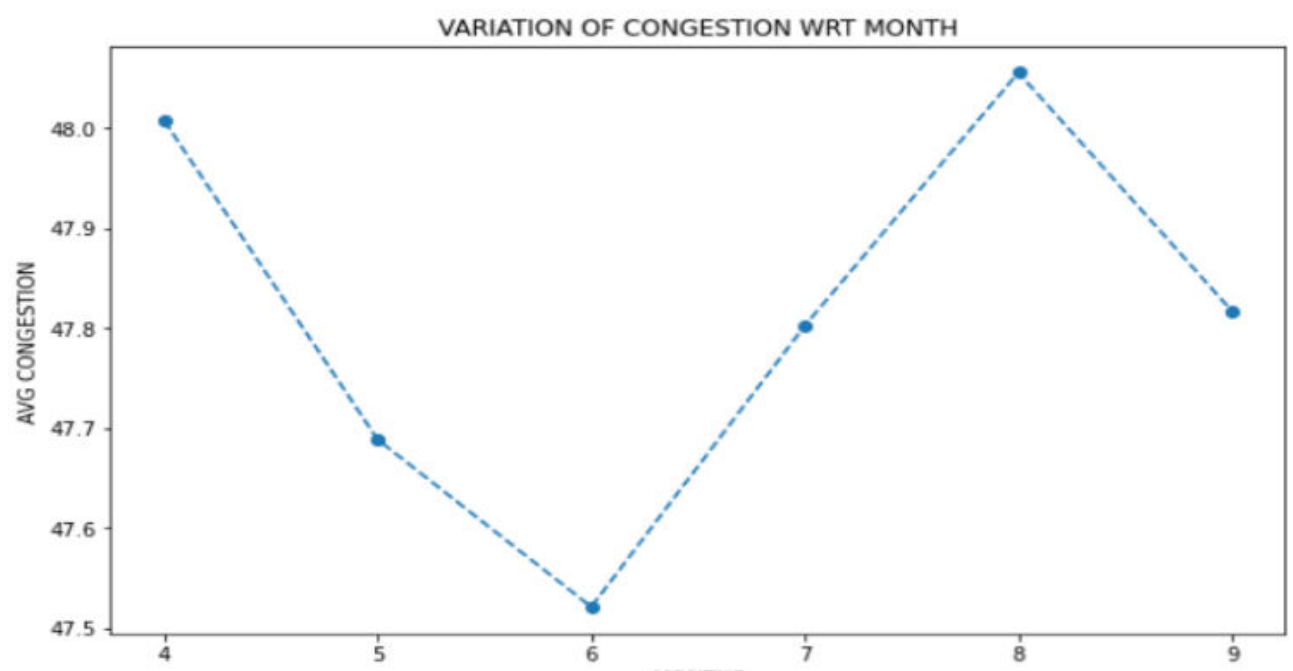

Fig. 4. Variation of congestion with respect to month

We can build upon the previous plot by splitting the average congestion based on two features. The first feature is the 'PM' in the dataset and the second feature is the 'Weekday' column. Fig. 5. compares the average congestion between weekdays and weekends and between AM and PM. Fig. 5(ii) shows that the average congestion was less during the weekends, and it was higher during the weekdays maybe indicating that in 1991 Chicago always had a busy week. Fig. 5(i) shows that as the day progressed the average congestion kept on increasing.

From the above and below figures we can infer very little with 'month' as the unit of measurement. Choosing a proper unit of measurement is often very important when it comes to Time Series Analysis.

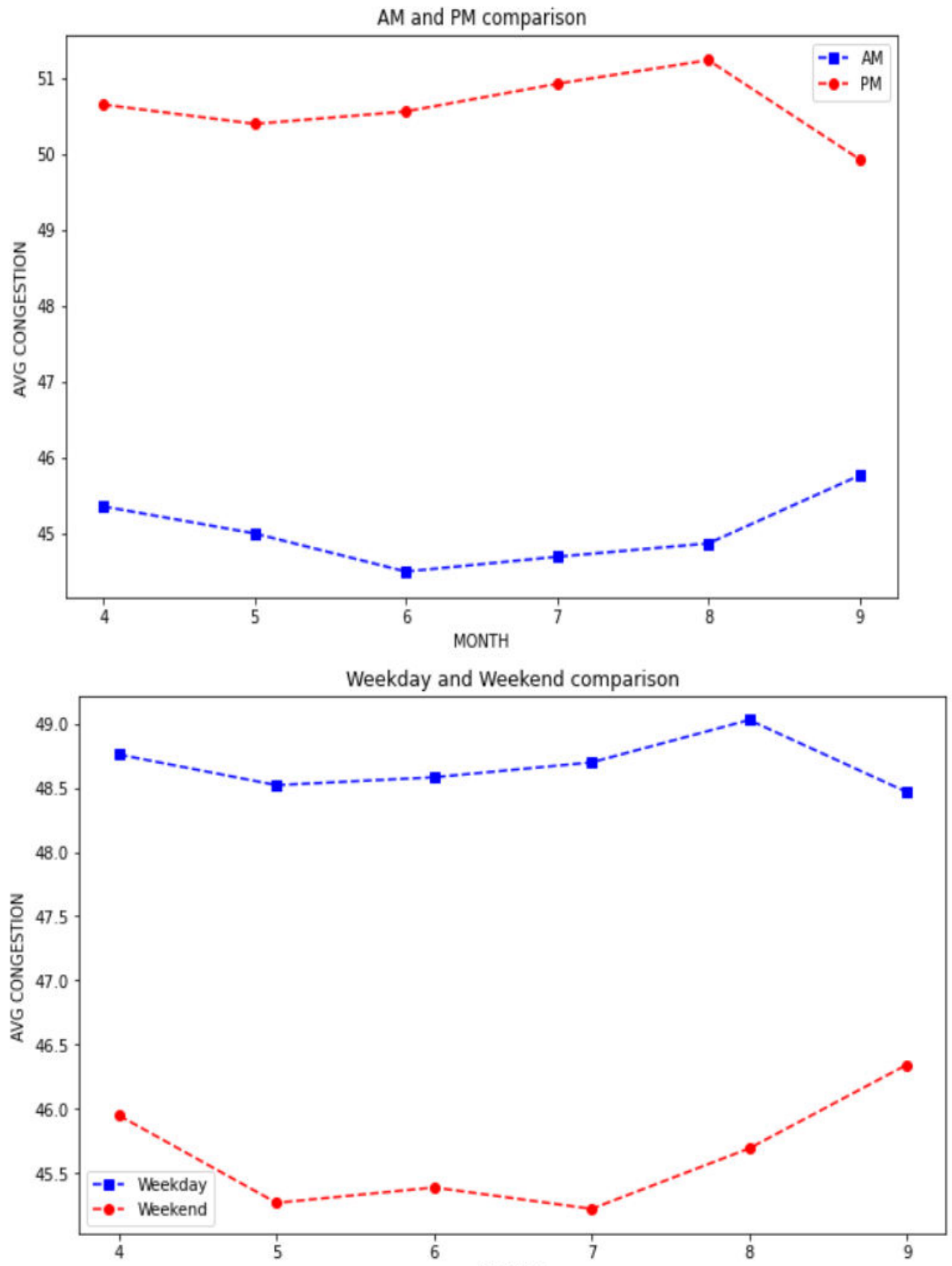

Fig. 5. (i & ii) Exploring the variations by using additional features

Since 'month' is not the best unit of measurement, 'hour' is used as the next unit of measurement which is shown in Fig. 6. Fig. 6. and Fig. 5(i) confirm that congestion increases as the day progresses and it starts to fall off around 6.00 PM. In general, whenever you are trying to measure a quantity like traffic congestion against some unit of time, 'hour' is considered as to be the best unit as it lies in the middle of both extremes like days and minutes.

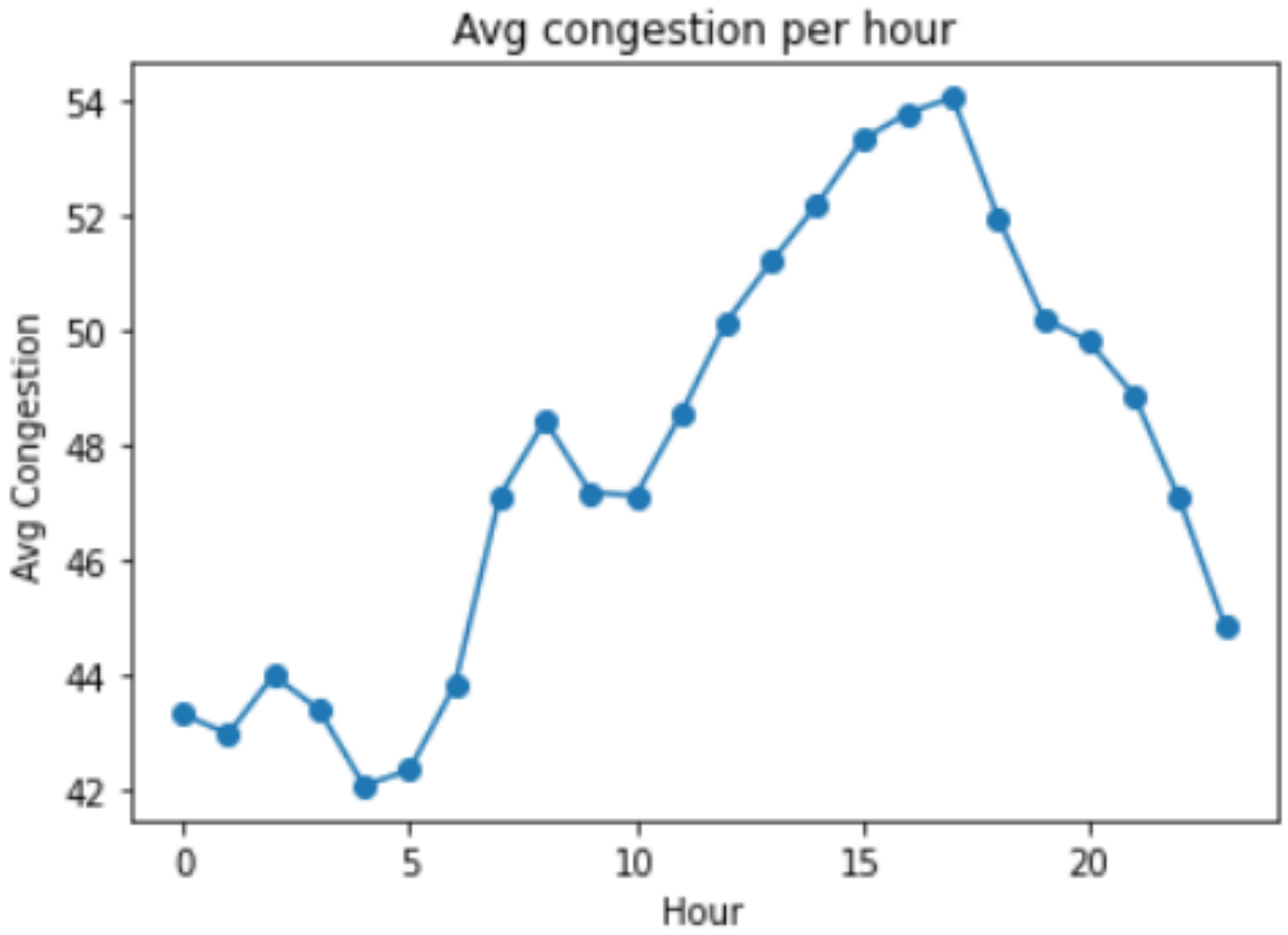

Fig. 6. Variation of congestion with respect to hour

Just like how Fig. 5. was developed from Fig. 4., Fig. 7. is developed from the previous figure. In Fig. 6. the average congestion with respect to hour has been plotted but this does not show us what is the variation for each day. Fig. 7. splits Fig. 6. into the seven days of the week and the plot gives us some interesting insights into the congestion statistics of Chicago. From the plot, you can see that all the days start with the same congestion but as the day progresses weekdays have a sudden spike in their congestion values. While Saturday and Sunday also follow a similar pattern their values are far less than those of the weekdays.

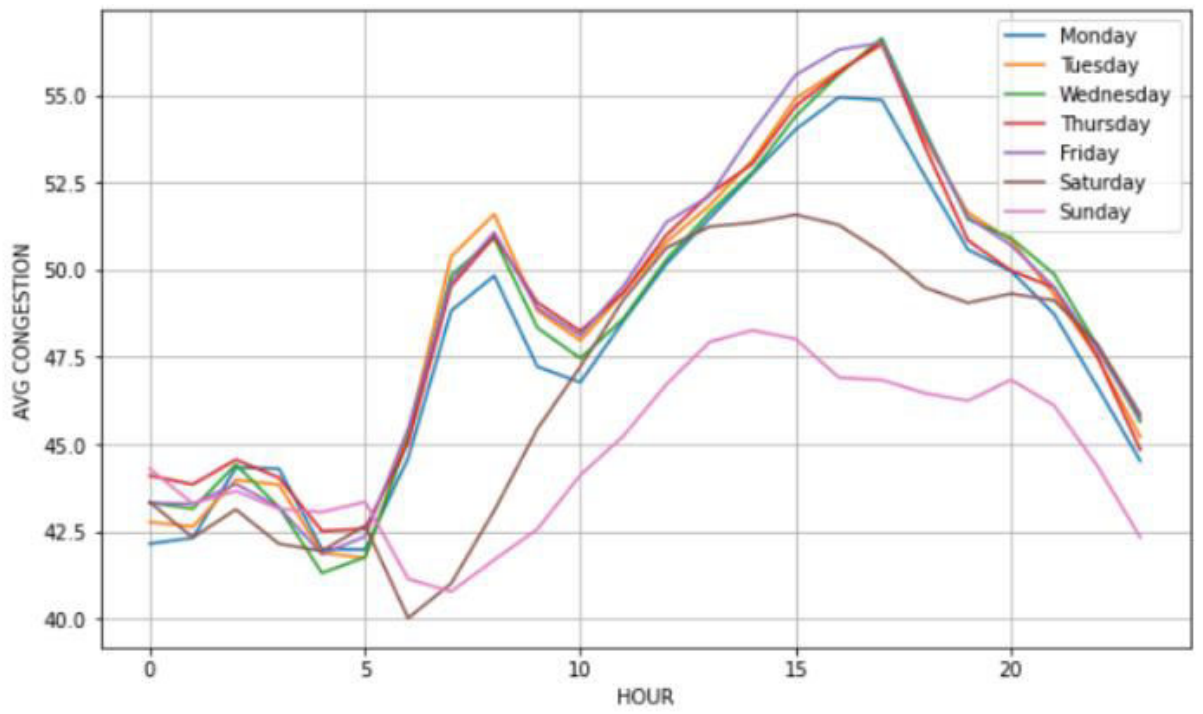

Fig. 7. Adding weekdays as 'hue' parameter

Fig. 8. uses the direction of travel to split the variation of congestion with respect to month and in this figure, there is a distinct separation between the congestion values for each direction. These sort of plots which separate out the target variable distinctly are very helpful for time series analysis as they enable us to confidently forecast values.

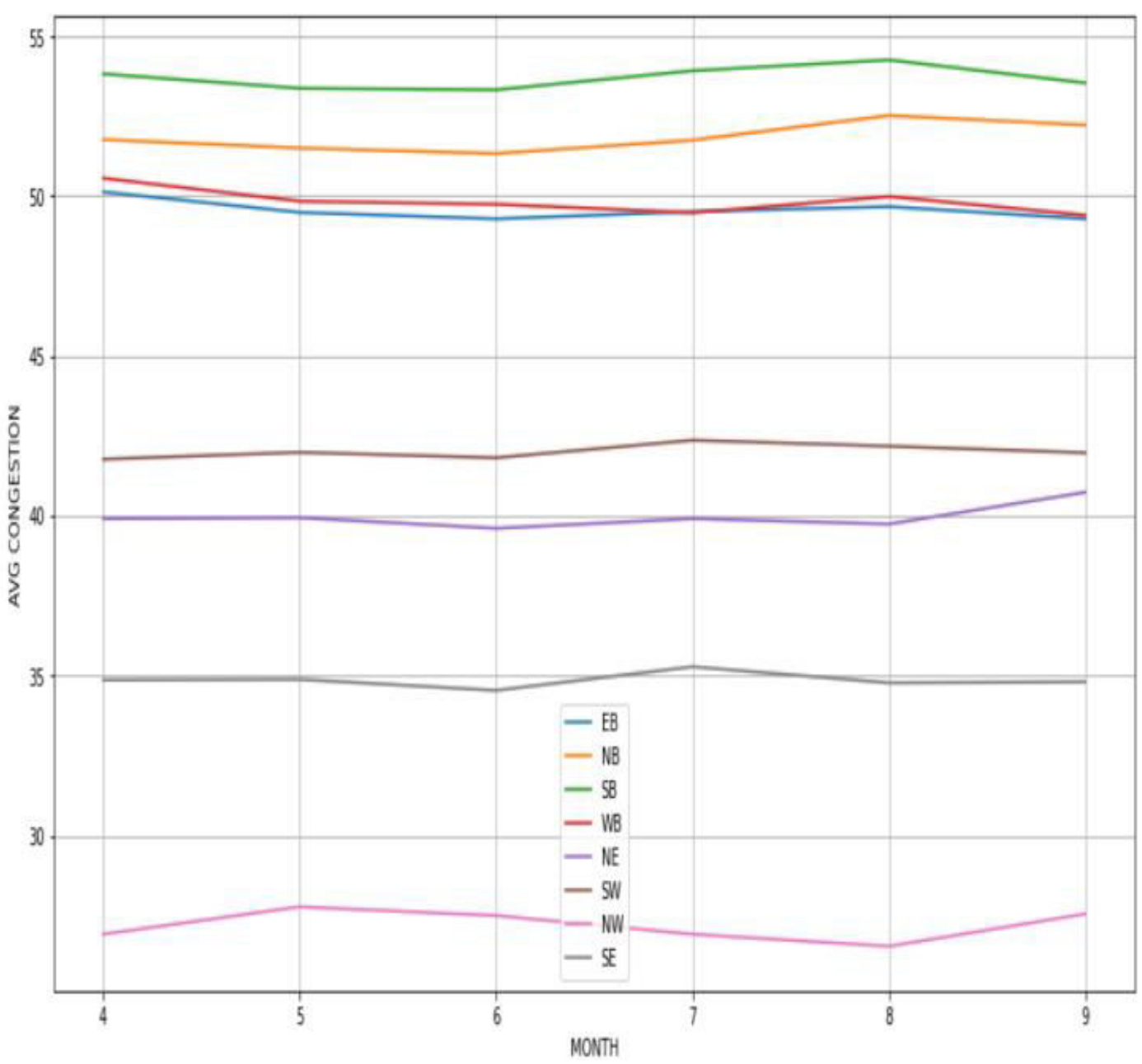

Fig. 8. Adding 'direction' as hue to the variation with respect to month

With the possible combinations of the co-ordinate features 'x' and 'y', a new dataframe is formed with 'month' as row and the combinations of 'x' and 'y' as columns. Features 'x' and 'y' by themselves won't contribute much to our understanding of the variation of congestion with respect to co-ordinates and therefore their combinations have been used. Each entry in the dataframe represents the average congestion corresponding to that particular month and that particular co-ordinate. Fig. 9. can be considered as an important plot as it can be used to predict which co-ordinates will require more surveillance and which do not, thereby optimizing the allocation of resources. Figure-9 consists of a lot of information, and it can be further decomposed to infer more information about each co-ordinate. Only four of the subplots have been shown here.

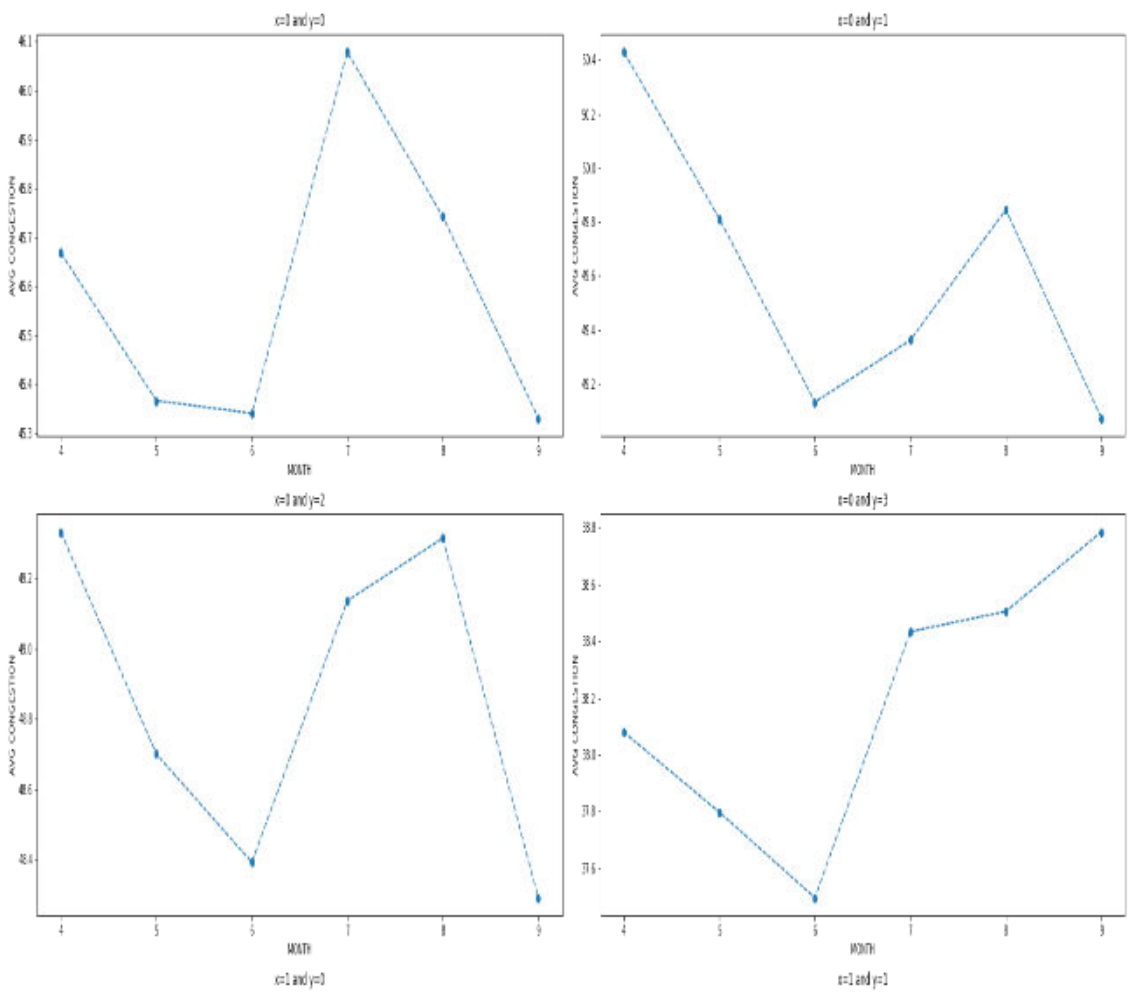

Fig. 9. Exploring variations with respect to month and co-ordinates

And finally, Fig. 10. plots out variation of congestion for each day of a month. Analysing the trend of a single day will do no good but analysing them as groups (for example, first 10 days of the month, last 10 days of the month, etc) will give us a lot of information. These types of plots require a lot of time to analyse as a lot of information is squeezed into a single plot. So, often these types of plots are decomposed into groups mentioned above and they are analysed separately.

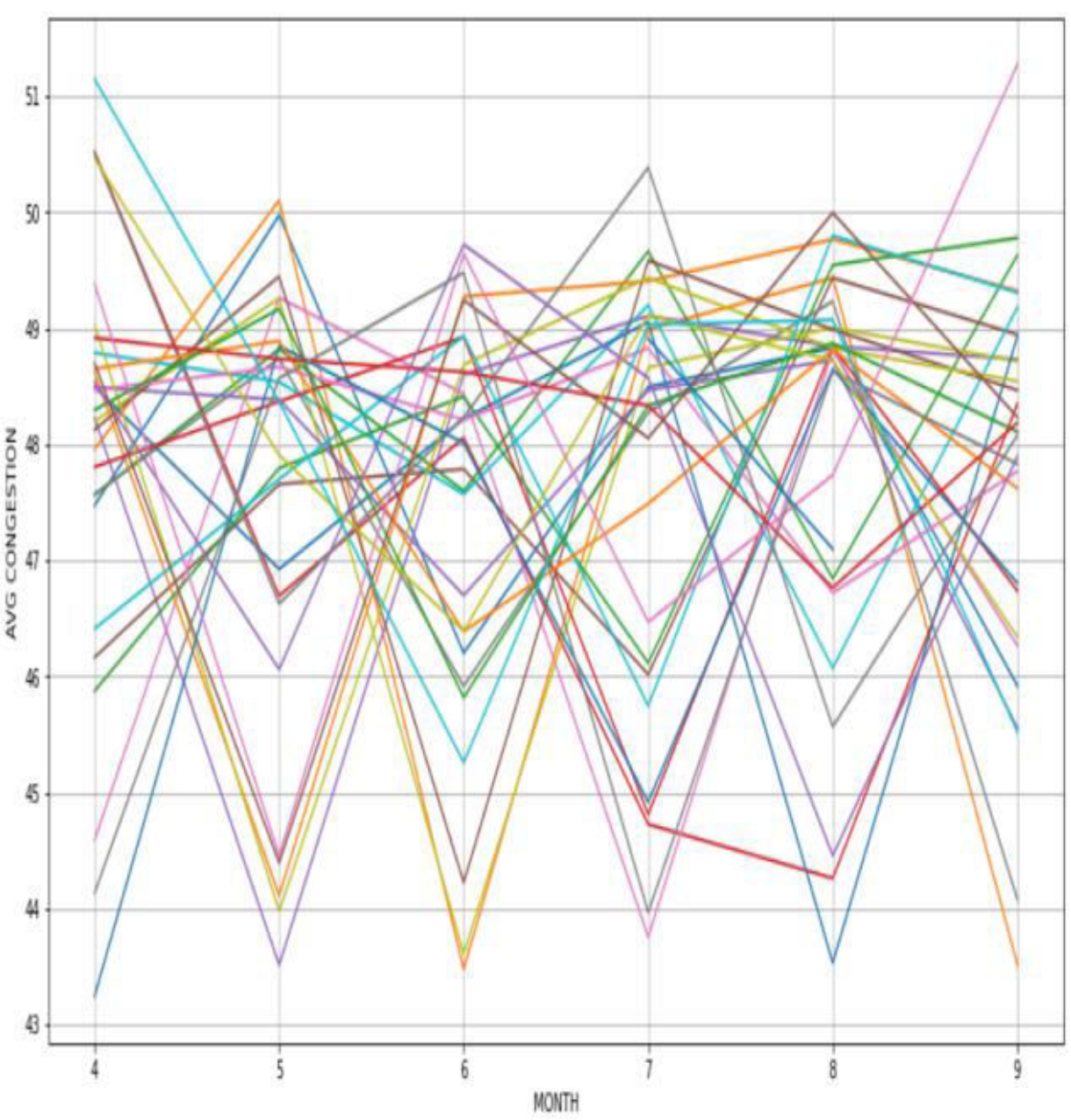

Fig. 10. Exploring variation with respect to month by adding 'day' as hue

## VI. CONCLUSION

There are a lot of ways to visualize a time series data and in this work, we have tried to provide a method that can be applied to most of the time series datasets. This work does not cover all the parts that form a time series rather it develops a method that can be used for all the constituent parts of a time series. Time Series is one of the areas where visualization can play a huge rule as the understanding of time series is highly visuals dependent. Just with visualization important trends and features can be figured out as the organization of time series data is like that.